\documentclass[preprint,journal]{vgtc}       

\ifpdf
  \pdfoutput=1\relax                   
  \usepackage{graphicx}                
  \DeclareGraphicsExtensions{.pdf,.png,.jpg,.jpeg} 
\else
  \usepackage{graphicx}                
  \DeclareGraphicsExtensions{.eps}     
\fi%

\graphicspath{{figures/}{pictures/}{images/}{./}} 

\usepackage{microtype}                 
\PassOptionsToPackage{warn}{textcomp}  
\usepackage{textcomp}                  
\usepackage{mathptmx}                  
\usepackage{times}                     
\usepackage{cite}                      
\usepackage{tabu}                      
\usepackage{booktabs}                  
\usepackage{tabularx}		
\usepackage{lscape}

\usepackage[table]{xcolor}
\usepackage{makecell}

\usepackage{xcolor}

\newcommand{\nc}[1]{#1}

\definecolor{orangeER}{RGB}{	204, 112, 0}
\definecolor{blueCT}{RGB}{24, 113, 201}

\usepackage{calc}
\newlength\myheight
\newlength\mydepth
\settototalheight\myheight{Xygp}
\newcommand*\inlinegraphics[1]{%
  \settototalheight\myheight{Xygp}%
  \settodepth\mydepth{Xygp}%
  \raisebox{-\mydepth}{\includegraphics[height=.85\myheight]{#1}}%
}

\newcommand{\Bar}{\texttt{Barchart1M}}
\newcommand{\Dor}{\texttt{DorlingSM}}
\newcommand{\Sym}{\texttt{GlyphSM}}

\newcommand{\tinyGeoLoc}{\raisebox{.3\height}{\inlinegraphics{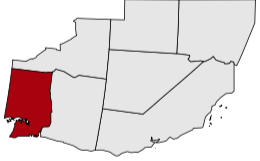}}}
\newcommand{\tinyGeoReg}{\raisebox{.3\height}{\inlinegraphics{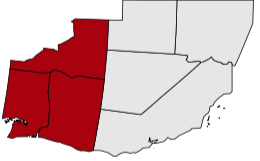}}}
\newcommand{\tinyGeoAll}{\raisebox{.3\height}{\inlinegraphics{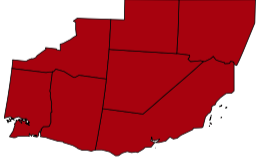}}}
\newcommand{\starEvidence}{\raisebox{.3\height}{\inlinegraphics{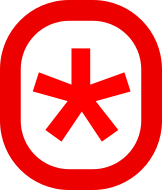}}}

\newcommand{\vGeoL}{\tinyGeoLoc \textsc{OneLocation}}
\newcommand{\vGeoR}{\tinyGeoReg \textsc{Region}}
\newcommand{\vGeoA}{\tinyGeoAll \textsc{AllLocations}}

\newcommand{\CT}{{Completion Time}}
\newcommand{\ER}{{Error Rate}}

\newcommand{\etal}[0]{{\em et al.}}
\newcommand{\ie}[0]{{\em i.e.}}
\newcommand{\eg}[0]{{\em e.g.}}

\onlineid{0}

\ieeedoi{10.1109/TVCG.2019.2934807}

\vgtccategory{Research}
\vgtcpapertype{please specify}

\title{A Comparison of Visualizations\\for Identifying Correlation over Space and Time}

\author{Vanessa Pe\~na-Araya, Emmanuel Pietriga, Anastasia Bezerianos }
\authorfooter{
\item
 Vanessa Pena-Araya is with Univ. Paris-Sud, CNRS, INRIA, Universit\'e Paris-Saclay. E-mail: vanessa.pena-araya@inria.fr.
\item
 Emmanuel Pietriga is with Univ. Paris-Sud, CNRS, INRIA, Universit\'e Paris-Saclay. E-mail: emmanuel.pietriga@inria.fr.
\item
 Anastasia Bezerianos is with Univ. Paris-Sud, CNRS, INRIA, Universit\'e Paris-Saclay. E-mail: anastasia.bezerianos@lri.fr.
}

\shortauthortitle{Pena-Araya \MakeLowercase{\textit{et al.}}: A Comparison of Visualizations for Identifying Correlation Over Time and Space}

\abstract{
Observing the relationship between two or more variables over space and time is essential in many domains. For instance, looking, for different countries, at the evolution of both the life expectancy at birth and the fertility rate will give an overview of their demographics. The choice of visual representation for such multivariate data is key to enabling analysts to extract patterns and trends. Prior work has compared geo-temporal visualization techniques for a single
thematic variable that evolves over space and time, or for two variables at a specific point in time. But how effective visualization techniques are at communicating correlation between two variables that evolve over space and time remains to be investigated. We report on a study comparing three techniques that are representative of different strategies to visualize geo-temporal multivariate data: \nc{either juxtaposing all locations for a given time step, or juxtaposing all time steps for a given location; and encoding thematic attributes either using symbols overlaid on top of map features, or using visual channels of the map features themselves.}
Participants performed a series of tasks that required them to identify if two variables were correlated over time and if there was a pattern in their evolution. 
Tasks varied in granularity for both dimensions: time (all time steps, a subrange of steps, one step only) and space (all locations, locations in a subregion, one location only). Our results show that a visualization's effectiveness depends strongly on the task to be carried out. 
Based on these findings we present a set of design guidelines 
about geo-temporal visualization techniques for communicating correlation.
} 

\keywords{geo-temporal data, bivariate maps, correlation, controlled study, bar chart, Dorling cartogram, small multiples}

\CCScatlist{ 
 \CCScat{K.6.1}{Management of Computing and Information Systems}%
{Project and People Management}{Life Cycle};
 \CCScat{K.7.m}{The Computing Profession}{Miscellaneous}{Ethics}
}

\teaser{
  \centering
  \includegraphics[width=\linewidth]{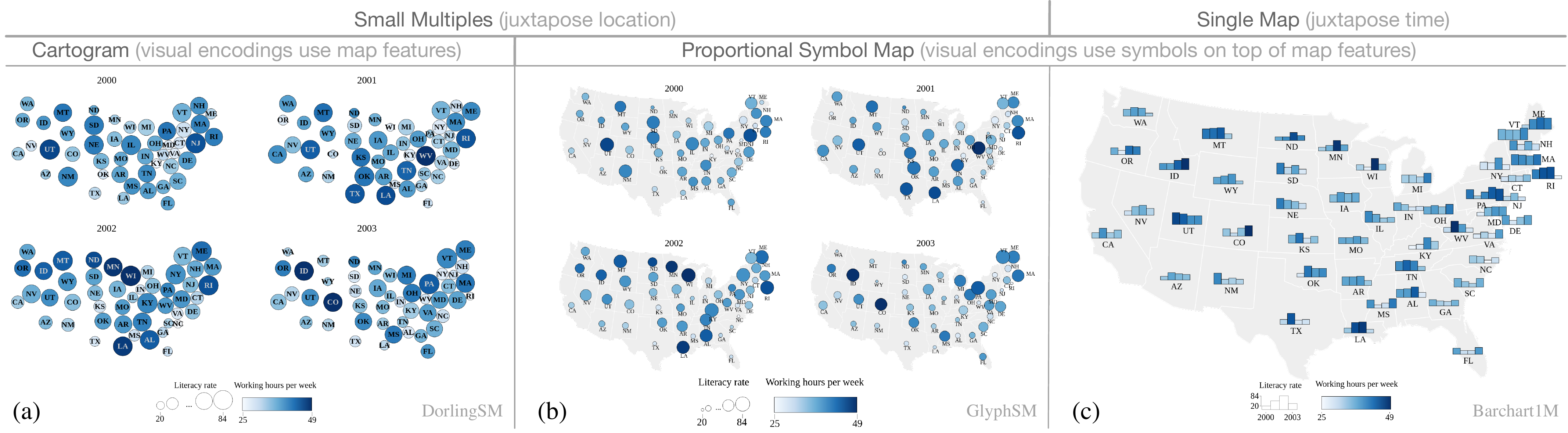}
  \vspace{-20pt}
  \caption{The three visualizations compared in our study. (a) Dorling 
  cartograms as small multiples, (b) proportional symbols (circles) on maps as small multiples, and (c) proportional symbols (bar charts) on a single map.
  In this example, each map shows the values of two artificially-created variables over four
  years. In each case, both variables have an overall positive correlation (Pearson 
  correlation coefficient $\geq$ 0.75) and no monotonic evolution.
  }
  \label{fig:teaser}
}

\vgtcinsertpkg

\begin{document}


\firstsection{Introduction}

\maketitle

Understanding phenomena often requires looking at multiple variables, their inter-relationships, and how these evolve over time. 
Take Hans Rosling's visualization of the demographics of countries in his seminal 2006 TED talk~\cite{Rosling:2006}. Looking at the life expectancy and the fertility rate together is key to understanding the phenomenon at hand. Watching their co-evolution provides many of the insights unveiled by the speaker.

In many cases, the data will also feature a spatial dimension. Rosling refers to individual countries, but also different groups of countries multiple times. The spatial dimension plays an important role in his story, even if it is only indirectly represented in the scatterplot. Again, understanding the interplay between the considered variables, and the spatial arrangement of the entities they describe, can yield key insights. 

This famous example illustrates the potential of multivariate geo-temporal data visualization as a storytelling device. The speaker communicates insights about two variables that are related thematically, and that describe a phenomenon that is situated both spatially and temporally~\cite{an2015space}. Beyond data storytelling, geo-temporal visualization can also support the analysis of such phenomena. The context, however, is different. While animation can 
illustrate temporal evolution when telling a story, it will often not be as effective for analysis purposes~\cite{Tversky:2002}. Moreover, depending on the application domain considered, information about group membership (\eg{}, a country belonging to a particular continent) might not be sufficient to understand what role the spatial dimension plays in the phenomenon. Thus more detailed information about 
 the topological relationship between entities might be necessary.

The problem of designing an effective visual representation in this context is challenging, as multiple data of different nature must be combined, each having specific characteristics: the thematic variables that describe the first-class entities in the dataset (life expectancy, fertility rate), the spatial properties of those entities (countries, continents), and the evolution of the thematic variables over time (years). 
Design choices will influence how well the representation can enable analysts to detect correlations between variables over space and time. It is thus important to identify guidelines to inform such designs. 

Prior studies have compared geo-temporal visualization techniques for a single variable that evolves over space and time~\cite{griffin2006comparison, Scheepens2014-qz, Livingston2011-ou, Livingston2012-yt}. Others have looked at two variables on a map (\nc{bivariate maps}), but at a specific point in time~\cite{Elmer2013-vw, Gao2018-so, Nusrat2018-qa}; 
or at how to visualize the correlation between two variables~\cite{Rensink2010-vg, Rensink2017-js, Kay2016-vx, Yang2019-fb}, including visualizations that can be used to depict temporal evolution~\cite{Harrison2014-ja}, but not in a geospatial context. To our knowledge, how effective visualization techniques are at communicating correlation between two thematic variables, that evolve over both space and time, remains to be studied.

\nc{We identify the different strategies used to combine thematic, spatial and temporal data into a visualization. 
The first design choice to be made  
concerns the combination of thematic variables in the representation: is the representation {\bf juxtaposing all locations for a given time step}; or {\bf juxtaposing all time steps for a given location}. The second choice concerns the visual encoding of thematic variables: either {\bf overlaying symbols on top of map features}; or {\bf using visual channels of the map features themselves}.}


\nc{We discuss design variations for each strategy and identify three candidate techniques (see Fig.~\ref{fig:teaser}).} Our study is designed to evaluate participants' ability to identify whether two variables are correlated over time or not, and if they are, if there is a pattern to their evolution.
As we expect the techniques to fare differently depending on the number of time steps and the number of geographical entities to consider, we test them on tasks that vary both in temporal and in geographical granularity.
Our results confirm this intuition, 
leading to a set of design guidelines about visualization choices for effectively communicating correlations in thematic geo-temporal data.

\section{Related work} \label{sec:relatedwork}

We 
first review some of the available visualizations categorized by how they
combine space and time, and then how thematic variables are encoded \nc{to 
create multivariate maps}.
We finally discuss research related to perception studies of visualizations of
correlated geo-temporal data.

\subsection{Visualizing Combined Dimensions of Space and Time}
Maps are the most direct visual representation of geo-temporal
data.
When combining both dimensions of space and time, thematic variables can be 
displayed by either juxtaposing locations (\eg, small multiples of compact map representations); 
or juxtaposing time (\eg, glyphs that represent multiple time steps overlaid on locations on a single map). 

\textbf{Juxtaposing location.}
From this category, small multiples are the most popular technique.
For example, Johnson~\etal{}~\cite{johnson2015analyzing} use small multiples
to observe the correlation of Internet adoption with GDP and with 
population over the years.
Animation can also be considered as a technique that juxtaposes location on 
maps that are presented in a sequence.
Animation has been used 
to smooth the transition
between views~\cite{craig2014animated}, or combined with symbols
to depict change~\cite{7847429}.
%

\textbf{Juxtaposing time.}
The most common approach is 
to use glyphs in 
2D (\eg,~\cite{fuchs2004visualizing, Park2017-wu, Kim2013-gr, Slingsby2018-pj,andrienko2004interactive})
or 3D (\eg,~\cite{tominski20053d})
on top of a single map.
Additionally, the 3rd dimension has been used 
to juxtapose time over a map.
For example, Space Time-Cubes~\cite{kraak2003space} arrange time steps
on the z-axis, effectively piling up the maps that correspond to each one of them.
They have been used in several
applications, \eg{},~\cite{Gatalsky2004InteractiveAO,TGIS:TGIS1194,ijgi2030817}.

\subsection{Visually Encoding Thematic Variables}

\label{sec:vgd}
%
Visually encoding data on a map can be done using two main strategies: mapping thematic attributes to visual properties \nc{of the map features; or overlaying symbols (\eg, basic shapes such as circles, or glyphs such as pie charts and bar charts)} on top of a base map, which remains untouched. 
As stated by Elmer~\cite{Elmer2013-vw}, the number of possibilities to create bivariate or multivariate maps can range 
from dozens to hundreds \nc{(the declarative model of Jo~\etal{}~\cite{Jo:2019} for multiclass density maps shows numerous examples). Thus in this section we focus on those representations that are most commonly used or 
studied.}

\nc{\textbf{Encodings that use visual channels of the map features.}}
Choropleth maps are among the most popular in this category~~\cite{guo2006visualization, stewart2010illuminated,  McNabb2018-wq}. \nc{They visually encode thematic attribute values using the map features' fill color.}
\nc{A bivariate type of choropleth}, called value-by-alpha maps, allows for two variables
to be displayed at the same time by combining color hue and transparency for 
each \nc{map feature}~\cite{Gao2018-so}.

Cartograms, 
use size as a visual encoding channel, 
and deform geographical shapes proportionally to the variable of interest~\cite{Nusrat2016-ix}.
There are four major types of cartograms:
contiguous, non-contiguous, Dorling and rectangular.
Contiguous cartograms distort regions to make their size reflect the thematic variable's value, preserving 
topology, and in particular adjacency, at the cost of statistical accuracy.
Non-contiguous cartograms rescale each region of the map independently.
They yield better statistical accuracy but fail to preserve topology (geographical regions are no longer contiguous).
Dorling cartograms~\cite{dorling1996area} yield more abstract representations of the geographical entities,
replacing each region with a circle (\autoref{fig:teaser}-a).
The circle's area can be mapped to a thematic variable.
The position of circles is computed so as to preserve the overall topology, putting each circle as close to its original 
location as possible, adjusting their actual position to avoid circles overlapping one another.
Finally, rectangular cartograms are similar to Dorling cartograms, but use rectangles
to represent each region, yielding even more abstract representations of the geographical entities.
Bivariate cartograms~\cite{tyner2014principles} use color or shade to encode a
second variable in addition to that mapped to size.
A recent variation on bivariate cartograms was presented by Nusrat \etal{}~\cite{Nusrat2018-qa},
in which two variables are visually encoded with size.

\nc{\textbf{Encodings that use visual channels of symbols on top of map features.}}
Overlaying thematic glyphs on top of a base map \nc{({\em ``symbols on maps"}~\cite{Harris:2000})} gives more flexibility 
compared to mapping data to the attributes of the \nc{map features} 
themselves. A wide variety of glyphs can be used to encode multivariate data.
They are typically placed on top of geographical regions, on an 
independent layer.
Proportional circles are the most frequently-used shape, but other
basic shapes like squares, triangles or any other symbol can also be used~\cite{tyner2014principles}.
Beyond simple shapes, more elaborate glyphs have been proposed; 
from generic glyph designs such as star glyphs or Chernoff faces~\cite{chernoff1973use}
to domain-specific ones such as those used in meteorology~\cite{Wittenbrink1996-ql}.

\subsection{Perception Studies on Correlated Geo-Temporal Data}


\begin{table}[]
\small
\begin{center}
\begin{tabular}{c|c|c||c}
                     & \makecell{Juxtapose \\ location} & \makecell{Juxtapose \\ time} & \makecell{No time} \\ \hline
\makecell{\nc{Visual encodings use}\\ \nc{symbols on top of map features}} & \makecell{\cite{Boyandin2012-th} \cite{Fish2011-sf} \\ 
 \cite{Scheepens2014-qz}* 
 } & \makecell{\cite{Kim2013-gr}$\dagger$
  \\ \cite{Livingston2011-ou, Livingston2012-yt}}  & 
\makecell{\cite{Elmer2013-vw}* \cite{Gao2018-so}*  \cite{Kaspar2011-rb}* \\ \cite{Sun2010-zt}$\dagger$ \\ \cite{Zhang2017-pf} \cite{Li_undated-ui} \cite{Yost_undated-ql} \cite{Beecham2017-bi} 
} \\ \hline
\makecell{\nc{Visual encodings use}\\ \nc{map features}} & \makecell{ \cite{griffin2006comparison} \cite{Nusrat2018-we}\\ \cite{Nusrat2018-qa}*}  & \makecell{\cite{Livingston2011-ou, Livingston2012-yt} \\ \cite{Nusrat2018-qa}*} & \makecell{\cite{Elmer2013-vw}* \cite{Gao2018-so}* \cite{Kaspar2011-rb}* \\ \cite{Sun2010-zt}$\dagger$ \\ \cite{Hagh-Shenas2007-wv}* \cite{Han2017-at}   \cite{mcnabb2018size} \cite{stewart2010illuminated} }\\   
\end{tabular}
\end{center}
\vspace{-10pt}
\caption{Categorization of 
studies 
comparing 
geo-spatial
visualizations. 
The first two columns represent the juxtaposition strategy. The third
groups studies which compare visualizations that do not include time. 
The two rows represent the categories of visual encodings (\nc{symbols or map features}). (*) indicates studies that consider more than one quantitative
variable, and ($\dagger$)  studies that consider one 
quantitative and one qualitative variable.
Note that some references are included in more than one 
cell as they make comparisons across categories.}
\label{table:studies_comparison}
\vspace{-15pt}
\end{table}

We now summarize the studies we consider most relevant to geospatial 
visualization. 
From the extensive literature, 
we selected a subset
using keyword searches
involving \textit{maps}, \textit{geographical}, \textit{geo-temporal}, 
\textit{empirical study}, \textit{evaluation}.
We filtered out papers that were more than 20 years
old, ones that consider numerical metrics but not visual perception (\eg,~\cite{Alam2015-qr, McNabb2018-wq}), or that
evaluated a new proposed technique in isolation (\eg,~\cite{Li2015-fx, Du2018-cb}).
The final set of 
articles can be
seen 
in \autoref{table:studies_comparison}.

We observe that most work on evaluating 
map-based visualizations does not focus on temporal evolution.
From the results of 
those that do,
we conclude that choosing the best-suited technique will depend on the task.
For example, for analyzing statistical data over time and space,
the results of Boyandin \etal{}~\cite{Boyandin2012-th} indicate that
users get more insights with small multiples than with animation.
This is confirmed by Robertson \etal{}~\cite{Robertson2008} for the analysis of 
trends using non geo-spatial visualizations.
For identifying moving patterns, Griffin \etal{}~\cite{griffin2006comparison}
show that animation leads to better results than small multiples.
Other studies that consider temporal change focus on comparing 
only two points in time (\eg,~\cite{Nusrat2018-qa, Nusrat2018-we}). They
do not provide insights about the compared techniques' performance for 
identifying trends over space and time.

Regarding visual encoding, we observe that most studies do not focus on more
than one quantitative variable at the same time. 
Particularly regarding correlation, two of them study
user performance for tasks that require analyzing the relationship between
two variables.
The first, from Gao \etal{}~\cite{Gao2018-so}, compares 
value-by-alpha maps with non-contiguous cartograms and proportional symbol
maps. The latter 
displayed 
better overall performance.
The second is from Elmer~\cite{Elmer2013-vw}, who evaluated eight different
visual encodings for bivariate maps.
He focused on studying the effectiveness of different combinations of visual variables for the analysis of patterns.
His results indicate that the eight combinations were consistent
in accuracy, showing the utility of bivariate maps. 
Time was not considered in these studies.

Other research studies the perception of spatial autocorrelation~\cite{klippel2011interpreting, Beecham2017-bi} (how much a phenomena is dependent on spatial location). Yet other studies investigate the perception of correlation in visualizations that do not involve maps~\cite{Rensink2010-vg, Rensink2017-js, Harrison2014-ja, 
Kay2016-vx, Yang2019-fb}. While such studies relate to our work, none of them considers all dimensions (correlation of {\em two} variables, over both space {\em and} time) simultaneously.

\section{Study rationale and Hypothesis} \label{sec:rationale}

The literature describes many visualization techniques capable of encoding two thematic variables in a geo-temporal context. As it would be impractical to test them all, we discard general strategies that are ill-suited to the context of visual analysis, and  identify representative techniques based on the strategies
briefly introduced earlier. We then motivate our tasks, formulate our hypotheses, and explain how we have generated the synthetic datasets used in the study.

\subsection{Selection of Visualization Techniques}

Our first decision is to discard techniques that use animation to convey the temporal evolution of thematic variables. There has been much discussion about the 
role of animations~\cite{Chevalier:2016} and their effectiveness~\cite{Tversky:2002}, 
with sometimes-contradictory findings. But there seems to be relatively broad consensus 
that they are ineffective for detailed analyses of multiple variables over sequences of many time steps: showing only a single step at a time, they require users to remember previously-seen steps, thereby increasing cognitive load~\cite{Harrower:2007}.

We also discard techniques that use 3D representations. These can provide more opportunities for mapping data attributes to visual variables (see, \eg{},~\cite{tominski20053d}), which can be useful when visualizing multivariate data. But they typically force users to interact more with the representation, and require more elaborate means of navigation because of the higher number of degrees of freedom, among other pitfalls~\cite{Shneiderman:2003}.

To make our study tractable, we make one final choice: to focus on 
visualizations based on how they represent the information, independently of any interaction technique. This means that we consider only static visualizations, in which elements can neither be filtered nor highlighted. As we discuss later in Sec.~\ref{sec:limfw}, follow-up studies should investigate how adding interaction impacts performance, but as this is the first empirical study to investigate the perception of correlation over space and time, there are already many factors to include before considering interaction techniques.


\nc{Based on these choices, we identify strategies} used to combine thematic, spatial and temporal data into one visual representation.
1) We first categorize visualizations according to how they organize thematic variables. They can juxtapose values for all locations at a given time step, yielding {\bf small-multiples} maps. Or they can juxtapose values for all time steps at a given location, yielding a {\bf single map}.
2) We then categorize visualizations according to how thematic variables are visually encoded~\nc{\cite{Elmer2013-vw}}. They can be mapped to the visual properties of \nc{symbols overlaid on top of the corresponding map features}, eventually forming a {\bf proportional symbol map}~\cite{Gao2018-so}. Or they can be mapped to the visual properties of the \nc{map features} themselves. Both choropleth maps and cartograms fall in this category, but we only consider {\bf cartograms} here. 
Indeed, encoding two thematic variables on choropleth maps is mostly limited to fill color hue, saturation and brightness, but these often interfere in terms of visual perception. 
Variations on the original design exist, such as, \eg{}, Banded Choropleth Maps~\cite{Du2018-cb}, but have not proven effective so far.

\nc{Combinations of these different strategies each yield multiple design variations. To avoid having to handle an unmanageable number of conditions, we choose at most one design per combination of strategies, and limit ourselves to designs that are actually used in practice.} Those choices are rationalized below, taking into account the fact that our two thematic variables are quantitative in nature.


{\bf Proportional Symbol Map + Small Multiples:} these techniques juxtapose values for all locations at a given time step. They consist of multiple identical base maps, one for each time step, with \nc{symbols superimposed on top of map features}. The symbols' visual channels encode the thematic variables, showing individual values for the corresponding time step. We select circles, as they are the most frequently used shape~\cite{tyner2014principles}, mapping the thematic variables to their radius and fill color brightness, respectively. This technique, which we refer to as \Sym{} in the study, is illustrated in Fig.~\ref{fig:teaser}-b.

{\bf Proportional Symbol Map + Single Map:} these techniques juxtapose values for all time steps at a given location. They consist of a single base map. Because all values for all time steps are juxtaposed, we can create \nc{miniature bar charts~\cite{Jo:2019}}, encoding one of the thematic variables using bar length instead of circle radius. Length is considered a more effective encoding channel than area, and this also makes for a more compact glyph than juxtaposed circles would. The second variable is mapped to each bar's fill color brightness. This technique, which we refer to as \Bar{}, is illustrated in Fig.~\ref{fig:teaser}-c.

{\bf Cartogram + Small Multiples:} these techniques juxtapose 
values for all locations at a given time step \nc{and encode thematic attributes directly on the map features, without using symbols}.
They consist of multiple cartograms, one for each time step in the dataset.
Among all variations on cartograms (discussed in Sec.~\ref{sec:vgd}), prior studies
have shown that contiguous cartograms and Dorling cartograms perform best 
overall~\cite{Nusrat2018-we}. 
We chose Dorling cartograms over contiguous cartograms as results 
of previous studies indicate they yield higher statistical accuracy and are
better suited to {\em summarize} tasks, therefore better aligned with the analysis 
of correlations.
This technique, which we refer to as \Dor{}, is illustrated in Fig.~\ref{fig:teaser}-a.

{\bf Cartogram + Single Map:} \nc{while instances of this combination do exist, all the ones we identified are somewhat contrived. 
Indeed, it is difficult to have a single small glyph meet all requirements: 
(i)~show two thematic variables; (ii)~show individual values for each of them, 
(iii)~for each time step; and (iv)~preserve the global topology of map features. 
One possibility would be to take the above Dorling cartogram, slice the
circles radially into as many time steps (transforming them 
into pie charts), and map the thematic variables to each slice's radius 
and fill color, effectively creating a rose chart. Such a design, however, makes it difficult to compare values across entities.
Other possibilities exist, involving, \eg{}, augmented donut charts or treemaps, but none of these is in reasonably widespread use 
and none stands out as a promising technique.
We thus did not include this combination in the study.
}

\subsection{Task Motivation} \label{sec:rationale_tasks}


Our goal was to compare the effectiveness of visualization techniques, when it comes to identifying 
the correlation between two 
variables and its evolution over time.
We had no hypothesis about which part is more difficult: detecting different types of correlations (positive / negative / non-existent), or characterizing their evolution (following a trend or not). 
We thus treat them as a single integrated task, that requires viewers to identify both potential correlations and their trends.  
\nc{We varied the combinations of these factors in our tasks to cover their range, but without exhaustively testing all combinations (Sec. \ref{sec:dataset_task}) and without making any assumption about their difficulty.}
Such integrated tasks fall under {\em ``characterize the relationship among multiple map features''} in Roth's task taxonomy~\cite{Roth2013-cf}.

To construct our tasks, 
we used the geo-temporal framework proposed by Peuquet~\cite{Peuquet1994-mr}, that describes
the linked triad of ``what'', ``where'' and ``when''.
Each task corresponds to a question of the type
\textit{when} $+$ 
\textit{where} $\rightarrow$ \textit{what},
where \textit{what} is the participant's characterization of the correlation and its evolution.

\begin{figure}
\includegraphics[width=\columnwidth]{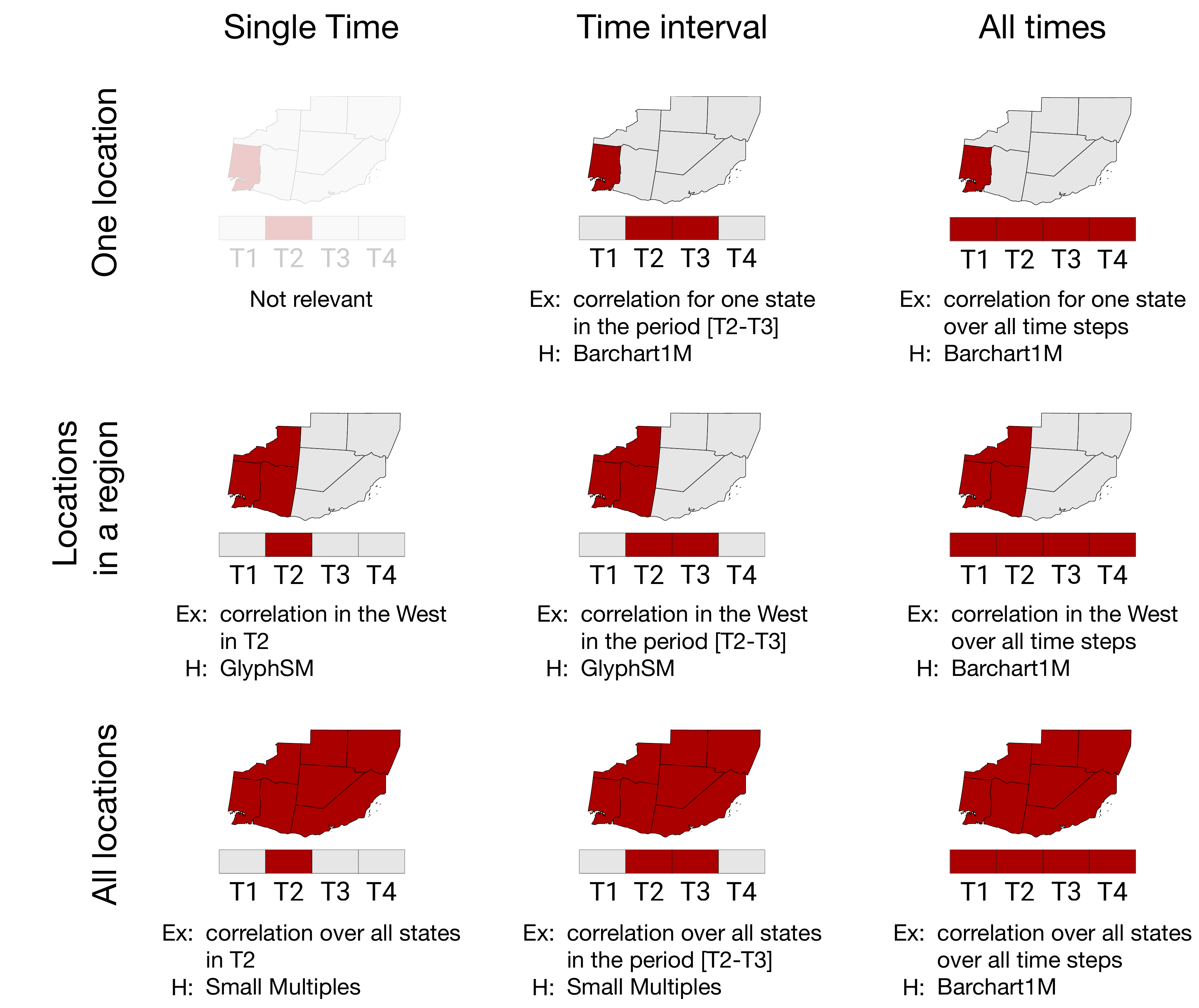}
\vspace{-10pt}
\caption{Summary of tasks based on spatio-temporal granularity. In each cell, the image illustrates the task,
together with an example (Ex), and our hypothesis (H) about which visualization will perform best overall.
}
\vspace{-10pt}
\label{fig:tasks}
\end{figure}

We varied the \textit{when} and \textit{where} in a way similar to other research (\eg,~\cite{Schiewe2018-pd, Goodwin2016-xv}),
using
three granularity levels.
In particular, the classification of granularity levels
for time (\textit{when}) is divided in (i) one time, (ii) a time interval, and (iii) all times.
Space (\textit{where}) is categorized as (i) one location, (ii) locations in a region, and (iii) all locations.
Crossing the spatial and temporal dimensions results
in a matrix of nine possible tasks illustrated in \autoref{fig:tasks}, together with a concrete example.
Correlation at one location in one point in time (top-left cell) is not meaningful 
and was discarded as a task. We thus ended up 
with eight possible spatio-temporal 
tasks.

We hypothesized that the best-performing visualization would depend on the task considered. Specific hypotheses are detailed in Sec.~\ref{sec:hypo}, and the techniques hypothesized to perform best for each task are also indicated in \autoref{fig:tasks}.


\subsection{Hypotheses}
\label{sec:hypo}

The following hypotheses capture our expectations and were  formulated before data was collected:

\textbf{H1:} We expect small multiples (\Sym{}, \Dor) to result
in better performance (less time and fewer errors) than single maps (\Bar) for tasks that require 
analysis at \emph{one point in time} only.
The search for the desired point in time is done only once across small multiples, 
and then the focus is on the spatial information that is grouped closely together.
Whereas for a single map the specific point in time needs to be identified
repeatedly across map features (bar charts).

\textbf{H2:} For  \emph{time intervals} in a single location, we expect better performance (time or errors) for a single map, as all information is colocated (one bar chart) (\textbf{H2.1}). When it comes to locations in a region, or to
all locations, small multiples (\Sym{}, \Dor) will fare better than 
single maps (\Bar) (\textbf{H2.2}). We expect 
that repeatedly identifying the right time interval across multiple 
locations in a single map 
will make this visualization slower and lead to more errors. 


\textbf{H3:} We expect single maps, that juxtapose time (\Bar), to result
in better performance  (time or errors) than small multiples (\Sym{}, \Dor) for tasks
that require analysis over all time steps. 
Indeed, small multiples require users to continuously 
change their focus between many maps to see trends for locations 
and make comparisons. 
This is not the case for single maps as they allow getting an overview of the 
behavior at each location quickly and identify trends.

\textbf{H4:} 
We expect that among small multiple techniques, \Sym{}, which overlays symbols on a base map, will feature better performance \nc{across all tasks}. \nc{Cartograms (\Dor) adjust the layout of features in each map independently, thus making it hard to identify and match them across small multiples.}

\subsection{Dataset and Task Construction} \label{sec:dataset_task}

For the setup of our experiment we use the map of the United States (\ie{}, 
map features are US states) over nine years of temporal evolution (\ie{}, a point 
in time is a year).

\nc{
The geography of the US states provides good diversity in terms of size of 
individual features (\eg{}, Texas compared to South Carolina) and density of those features
(\eg{}, west coast compared to east coast).}
%
%
\nc{In trials involving a single location, we varied the size of target features (smaller \& larger states) and density of the surrounding geographic area.}
We grouped locations in \nc{contiguous} regions using the four cardinal points:
north, south, west and east.
\nc{These regions were selected as they represent common geographic division of
countries or other administrative levels.
Regions were determined by drawing an imaginary line that divided the country
into two equally-sized areas, vertically for east and west, horizontally for
north and south.}
\nc{This resulted in areas of varying density across trials.
To avoid participants fixating their gaze over discontinuous areas,
especially for tasks involving a subset of locations in a region (\autoref{fig:tasks}, second row), we removed Alaska and Hawaii from the map.
This resulted in a total of 48 locations, a fair amount of 
locations to analyze.}


Regarding time granularity, \nc{we define all time spans to be nine years long (a number that utilizes the space of a small multiple setup)}. Time intervals were made of four consecutive 
steps, selected in the middle of the range so as not to favor single
maps -- identifying the first or last part of the small bar charts is much easier.
\nc{Four years represent almost half of the total time steps, which allows us to balance
the amount of patterns (correlations to identify) and noise (additional data-points to make the task realistic).}

The two variables were presented to participants as {\em literacy rate} and {\em working
hours per week}.
Nevertheless, to control the displayed correlation and trends within the different  
spatio-temporal constraints, we used artificially-created datasets.
We initially created variables that followed normal distributions, as other
perception studies about the visualization of correlation do~\cite{Rensink2010-vg, Harrison2014-ja}.
With this type of distributions, it is common that points do not follow
strict patterns of both increasing at the same time (in case of positive correlation),
or one increasing as the other decreases (in case of negative correlation).
This is not a problem with scatterplots, as the overall 
distribution of many points helps convey the overall relationship. 
However, in our case, the number of points in time was small, minimum 4 for 
time intervals and maximum 9 for all time steps.
Thus, even if one point did not follow the pattern, it would suggest that there was no correlation.
We instead generated pairs of points using a random linear regression model with 
added Gaussian-centered noise,\footnote{Data was created with 
Scikit-learn~\cite{scikit:2011}, using \texttt{make\_regression}.} as
\nc{the difference between values could be evaluated more clearly.} 
The obtained points were checked to ensure that they follow the pattern for the desired time range.
To make the generated distributions 
closer to 
actual literacy rate and working hours per week, 
we scaled our generated data  
between values extracted from Rosling's GapMinder example.
For instance, for the variable assigned to literacy rate, we scaled between 
a minimum within $[20, 30]$ and a maximum within $[75, 85]$.
For the variable assigned to working hours per week, the minimum varied within
$[25, 35]$ and the maximum within $[40, 50]$.

For each task, we created a dataset that followed particular spatio-temporal patterns. 
The possible correlation patterns were: positive correlation ($r \geq 0.75$) with and 
without monotonic evolution; negative correlation ($r \leq -0.75$) with and
without monotonic evolution; and no correlation ($|r| \leq 0.2$). 
These patterns were enforced for the space and time granularities considered in each task (\eg{}, a time range or all times).\footnote{We note that for tasks that require analysis in one point in
time, it was not relevant to create two variables with monotonic evolution.}
We added distractors 
for the locations and time points that were not the focus of the task 
by including 1/3 of data points that did not follow the assigned pattern. 

To increase reliability, our design included three repetitions per task, that were aggregated in our analysis.
To avoid learning for each repetition, we varied the selected location, region, point 
in time and time interval.
We generated one dataset per task repetition that, for
the spatial and temporal constraints required by the question (time and space granularity),
followed a different correlation pattern.
For the three repetitions, there was always: one trial with no correlation, one with correlation (positive or negative) but no monotonic evolution, and one with correlation (positive or negative) and monotonic evolution. Fig.~\ref{fig:task_answers} illustrates the different configurations.

To avoid participants remembering answers across visualizations,
from each generated dataset 
we derived two 
additional 
variations 
by shuffling data
over states, over years, or both. 
Thus, for each task repetition displayed in each visualization, the 
participants would observe a dataset with 
the same 
structure but with different
layouts. 
In total, this resulted in 80 datasets: 8 tasks $\times$ 3 repetitions $\times$ 3 datasets (1 original + 2 shuffled variations)
$=$ 72 for main trials $+$ 8 for training. 

\section{Study design} \label{sec:design}

The study was designed to evaluate, for each of the tasks, the three visualizations introduced earlier.
\nc{Supplemental material containing dataset generation code, experiment data, analysis scripts and detailed results are available at {\small \url{http://ilda.saclay.inria.fr/spacetimecorr}}.}


\subsection{Experimental Design}\label{sec:exp-design}

\begin{figure}
\includegraphics[width=.9\columnwidth]{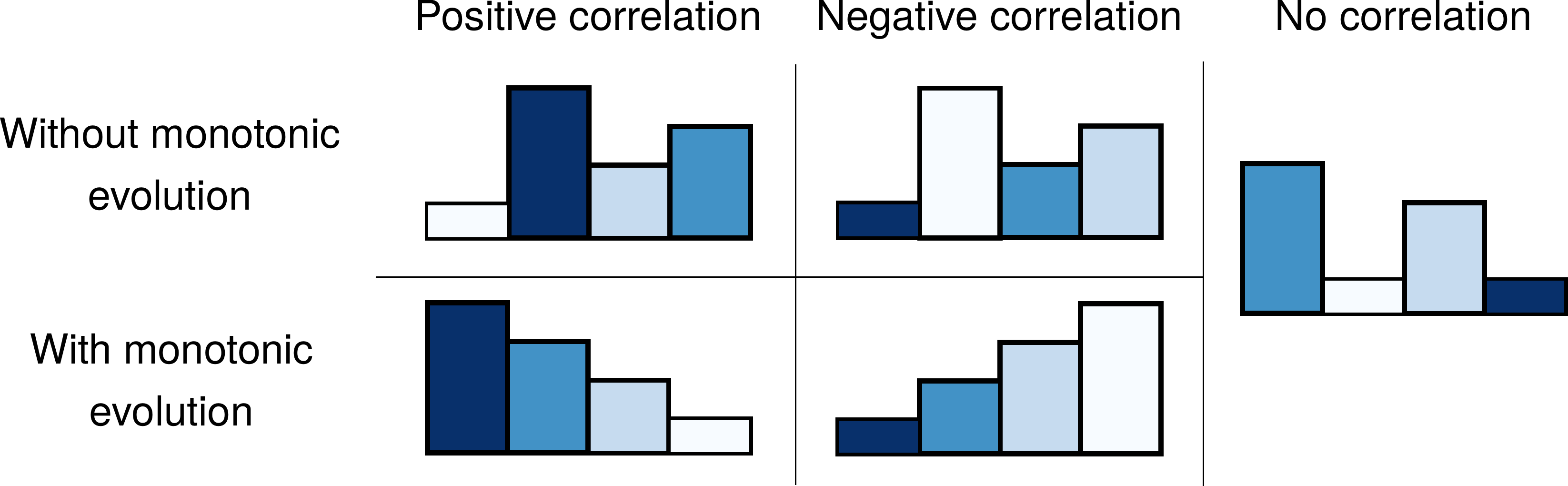}
\vspace{-5pt}
\caption{\nc{Schematic illustration of all possible answers for tasks in} \autoref{fig:tasks}. \nc{The three trial repetitions included combinations, such that each correlation type (positive, negative, no-correlation) appeared once. When temporal evolution was applicable, one of the positive/negative correlations was coupled with monotonic evolution, while the other was not.}}
\label{fig:task_answers}
\vspace{-10pt}
\end{figure}

We used a within-subjects design where all participants were exposed to 
all three visualization techniques.
For each technique, a participant had to perform 8 training trials, and 8 measured tasks
$\times$ 3 repetitions $=$ 24 main trials.
Repetitions considered one of each possible correlation types: 
Positive, Negative or No-Correlation. 
For tasks that involved analysis over time, answers also included monotonic choices \nc{(\autoref{fig:task_answers})}.

Technique presentation order and dataset variations were
counterbalanced across participants using Latin squares.
Tasks were grouped by time granularity (one point in time, time interval, all 
times) and their order of presentation was counterbalanced as well. 
For each time granularity, the order of geographical
granularity was randomized. 
Within each group of space and time granularity, the three task repetitions were also
randomized.
In total, the experiment consisted of 18 participants $\times$ 3 visualizations 
$\times$ 8 tasks $\times$ 3 repetitions $=$ 1296 trials.

\subsection{Apparatus and Participants}

We used a 27'' Apple Thunderbolt Display set to its default resolution (2560$\times$1440
pixels).
The web user interface was implemented in Django and visualizations were generated with
D3~\cite{Bostock:2011} and Vega~\cite{Vega:2016}.
We made sure that all visualizations were of similar size by keeping their width consistent (adjusting height to keep the original aspect ratio).
All visualizations fit comfortably on the screen and did not require scrolling.  
More specifically, the dimensions were 1350$\times$996 
pixels for \Sym{} and \Dor{}, and 1350$\times$849 pixels for \Bar.

We recruited 18 participants before starting the experiment, a number that allowed us to counterbalance technique presentation order. We continuously recruited participants until we arrived at this pre-defined 
number.
Our participant exclusion criteria included: not completing all conditions, or failing any of the 3 training trials. 
Given the complexity of the task, we assumed task learning would transfer across techniques. Thus, 
an excluded participant would have to be replaced with another participant with an 
equivalent configuration of technique, dataset and task presentation ordering.
We had to replace a single participant who declared during the second session
that she had misunderstood how to perform the tasks in the first session.

From the final 18 participants (10 female and 8 male), none reported 
any color deficiency. All had normal or corrected-to-normal vision. 
Age ranged from 23 to 40 ($M = 27.6$, $SD = 4.9$) and most of them were 
students (13/18) from either a PhD or a Masters' program.
Their backgrounds were mainly HCI, Computer Science and Visualization.
They were all volunteers, and did not receive any monetary compensation. 

\subsection{Procedure}

\begin{figure}[t]
\includegraphics[width= 1\linewidth]{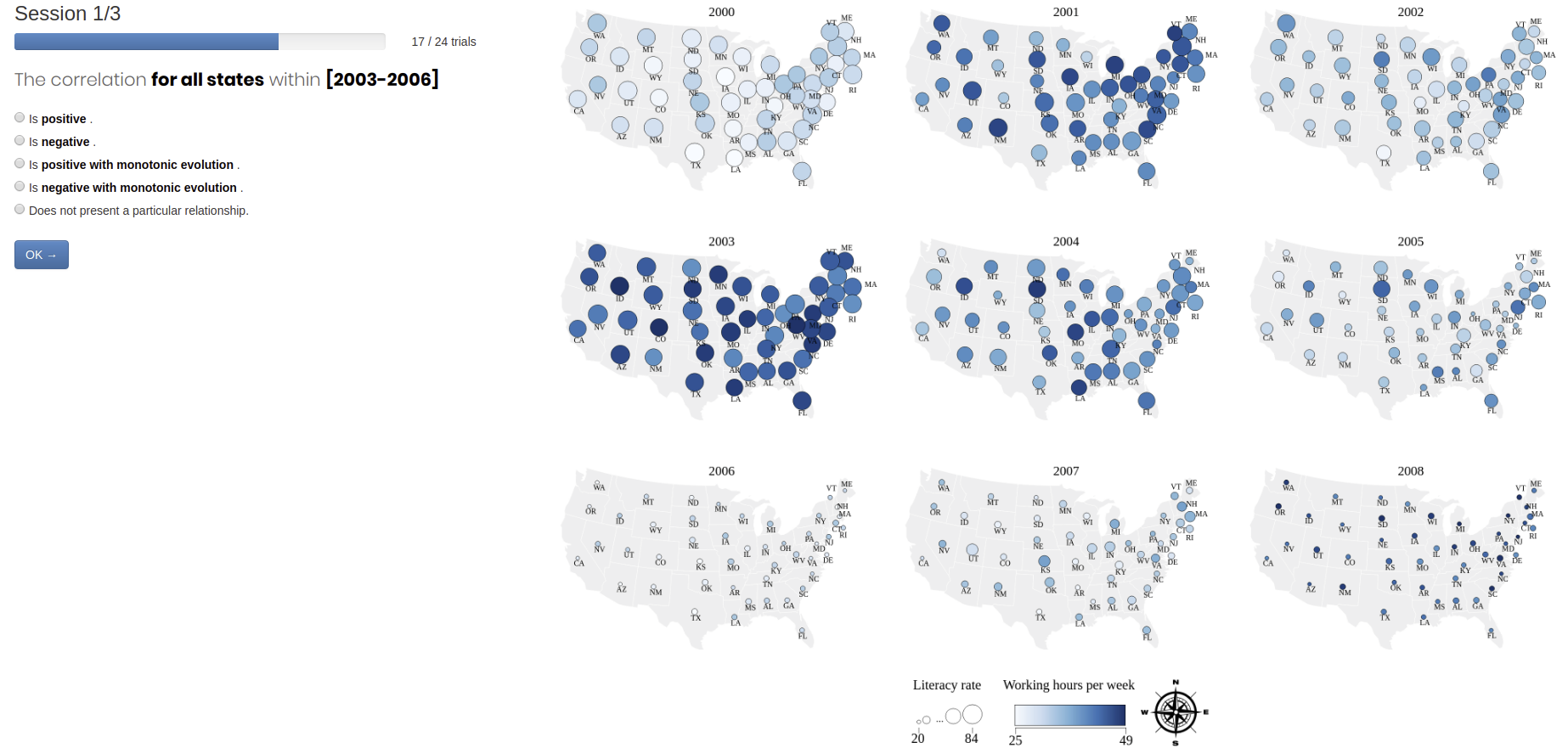} 
\caption{Web interface used to conduct the experiment. Visualization = \Sym{};
task performed on a {\sc Time Interval}, for {\sc All Locations}.}
\label{fig:study_interface}
\end{figure}

First pilots of our study showed that conducting the tasks was mentally 
demanding. We thus divided the study in three sessions, one per
visualization, performed on three different days \nc{(that could be
consecutive and at most 9 days apart)}.
Each session consisted of three parts: introduction, training, and main 
trials.
In the first session, participants signed a consent form, were told that they could withdraw at any time, and filled
out a demographic questionnaire. 

\noindent\textbf{1) Introduction and training.} 
The experimenter explained the visualization to be used in the 
session, along with examples of how correlation and monotonic evolution looked on
it.
Further training was conducted, that consisted in 
answering eight trials, one per task (described next).
After finishing each trial, the system would indicate if the answer was 
correct or not. If participants made no error and declared that they had no further question, they 
would start the main trials. Otherwise, 
the experimenter would add further training trials. 

\noindent\textbf{2) Completion of main trials.} 
%
%
\nc{Fig.~\ref{fig:study_interface} shows a trial screenshot. On the left are the overall progress, the question asked and possible answers. On the right is the generated visualization for that condition.
Before each trial a map was shown, highlighting the location(s) that the trial would be about. Our aim was to reduce potential bias due to prior knowledge of the United States' geography, and to ensure there was no ambiguity about geographical features to consider such as, \eg{}, which states constitute a region.}

Participants completed 24 main trials per session (visualization). In this phase, they did not get any feedback about the correctness of their answers. They were instead asked to report the level of confidence in the answer they had just given (low, medium, high).

Once trials were completed for a visualization, participants filled out a 
post-hoc questionnaire about the strategies used to complete the eight tasks, and how easy it was to complete each one of those tasks.
After finishing the third session, participants filled out a final questionnaire, in which they were asked
to rank the visualizations. \nc{A representative image of each visualization was displayed in the form to help participants remember them.}
The entire experiment (3 sessions) took approximately one hour and a half. 

\subsection{Measures}
For each task, we defined three metrics, two objective, one subjective:\\
\noindent
$~~$\texttt{{\small - Task completion time:}} measured from the moment
participants saw the trial screen until they submitted an answer. We computed the average over the 3 repetitions. \\
$~~$\texttt{{\small - Error rate:}} computed as the number of incorrect answers per task multiplied by the 
total number of repetitions.\\
$~~$\texttt{{\small - Self-reported confidence:}} measured on a 3-point Likert scale (high, 
medium, low).

For each technique, we recorded:\\
\noindent\texttt{{\small - Strategies to complete the trials:}} described as free text.\\
\noindent\texttt{{\small - Self-reported difficulty to complete each type of task:}} 
measured on a 5-point Likert scale from very easy (5) to very difficult (1).

\section{Results}
%
%


We analyze, report, and interpret all our inferential statistics using graphically-reported point estimates and interval estimates \cite{Cumming05inferenceby,Dragicevic:2016:FSC}. 

We report sample means for \textbf{\CT} and \textbf{\ER} and 95\% confidence intervals (CIs), indicating the range of plausible values for the population mean. For our inferential analysis we use means of differences and their 95\% confidence intervals (CIs).\footnote{A CI of \emph{differences} that does not cross 0 provides evidence of differences - the further away from 0 and the smaller the CI the stronger the evidence.}
We use BCa bootstrapping to construct all confidence intervals (10,000 iterations). 
Since in our \emph{per-task} analysis we test specific predictions rather than a universal null hypothesis, no correction for multiple comparisons was performed \cite{Cumming2014,Perneger1236}. 
A p-value approach of our technique can be obtained following the recommendations from Krzywinski and Altman~\cite{Krzywinski:2013:PoS}.
%
Finally, we also report percentages for self-reported \textbf{Confidence} results. 

We analyzed a total of 1296 trials (18 participants $\times$ 72 trials). 
All reported analyses were planned before the experiment started. 

We first provide an overview across tasks.\footnote{\nc{We counterbalanced visualization order across participants to mitigate learning (Sec.~\ref{sec:exp-design}). An unplanned analysis indicates that although participants improved over sessions (performed best in the 3rd visualization presented than in the 1st), there was indeed no evidence of asymmetric learning across \Bar\ and \Sym, thus counterbalancing worked for them (there is some Time improvement for \Dor).
Analyses/charts are available as supplementary material.} 
}
Since our hypotheses are task dependent, we then perform a detailed per-task analysis.

\subsection{Overall results across tasks}

\begin{figure}[h]
 \vspace{-10pt}
    \includegraphics[width= 0.49\linewidth]{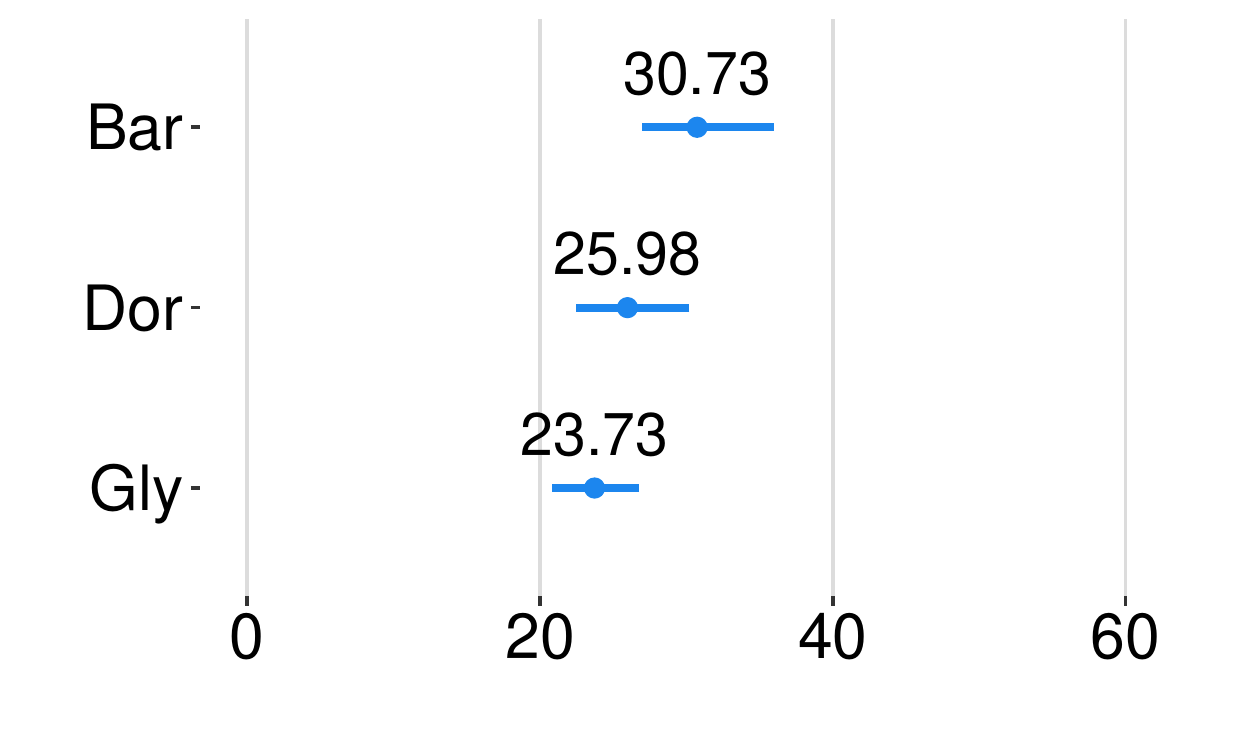} 
    \includegraphics[width= 0.49\linewidth]{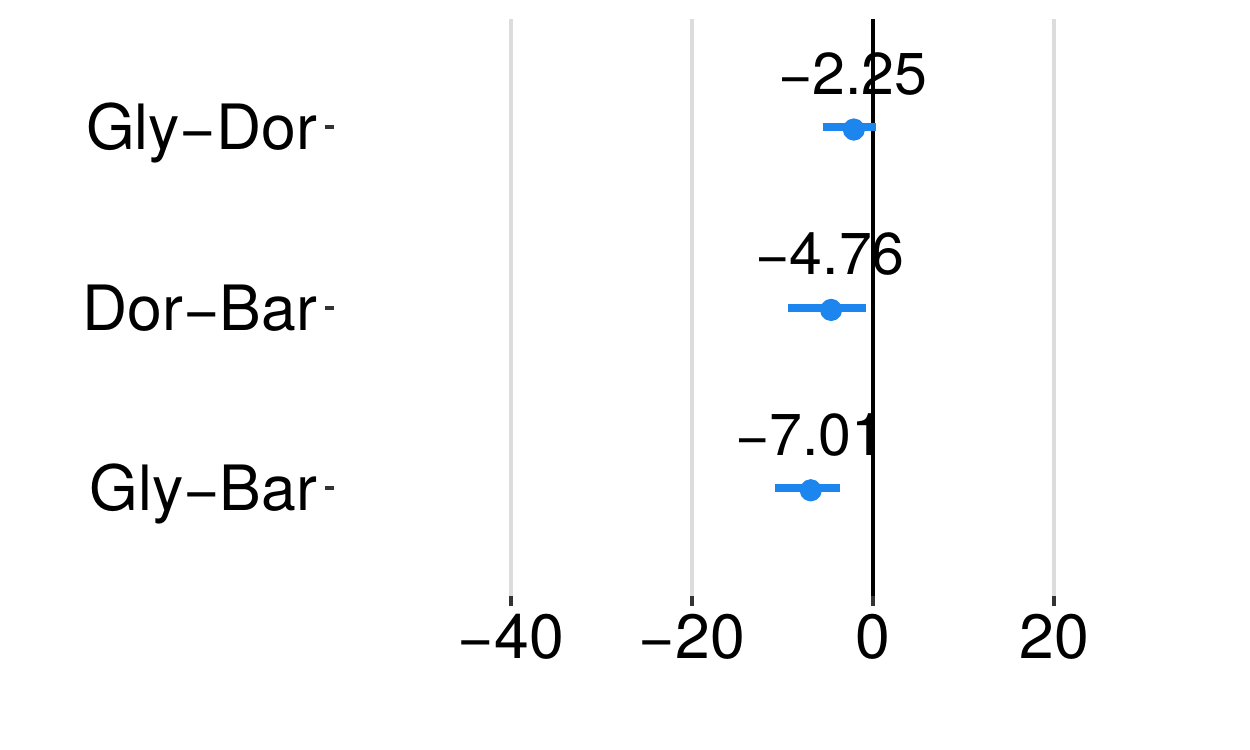}     
        \vspace{-20pt}
     \caption{Left: Mean \CT\ in seconds  for each visualization, for all tasks. Right: Pairwise comparisons for each visualization. Error bars represent 95\% Bootstrap confidence intervals.
     }\label{fig:all_time}
\end{figure}

\noindent \textbf{\CT :} \autoref{fig:all_time} shows completion times of all tasks collectively. Mean times per technique are on the left, mean differences on the right. Mean times are shorter for \Sym\ (23.7sec) followed by \Dor\ (26sec) and \Bar\ (30.7sec). 
There is strong evidence that \Bar\ is slower than \Sym\ (by 7.0sec on average) and evidence that it is also slower than \Dor{}, although the difference is smaller (4.5sec on average).

\begin{figure}[h]
    \includegraphics[width=.49\linewidth]{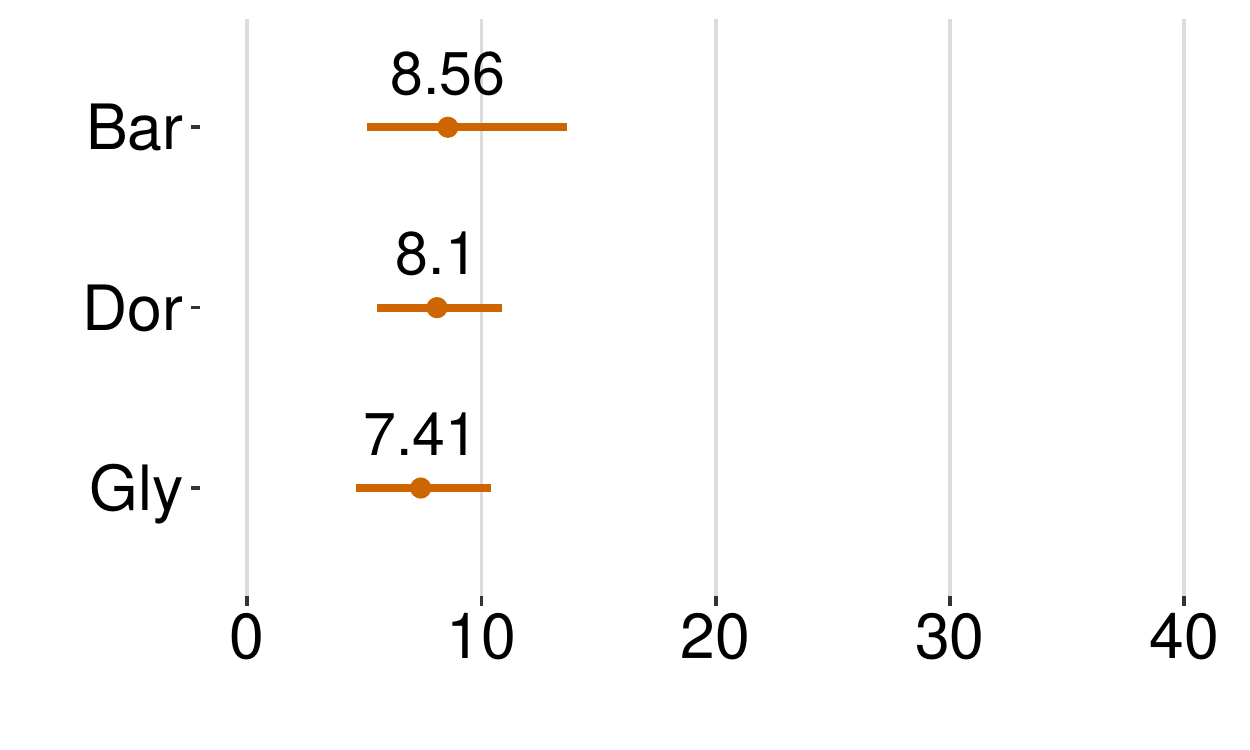} 
    \includegraphics[width=.49\linewidth]{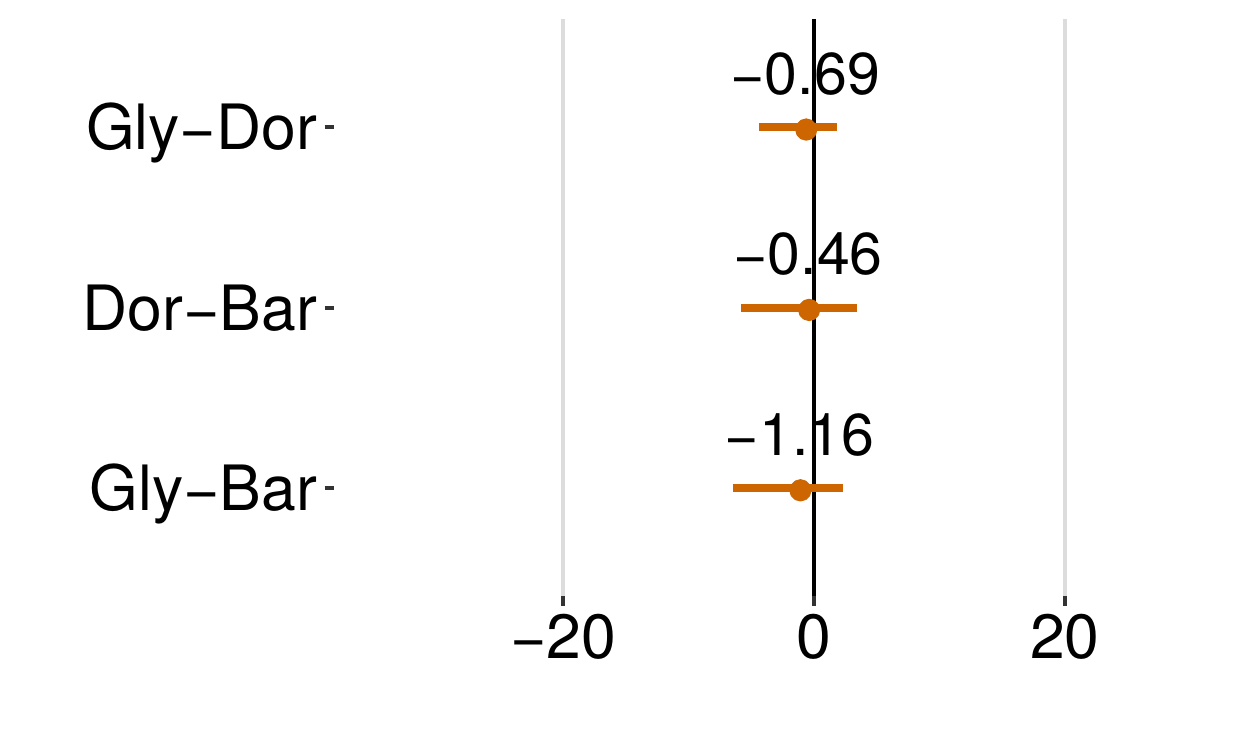}     
    \vspace{-20pt}
     \caption{Left: Mean \ER\ in \% for each visualization, for all tasks. Right: Pairwise comparisons for each visualization. Error bars represent 95\% Bootstrap confidence intervals. 
     }\label{fig:all_error}
\end{figure}

\noindent \textbf{\ER :}  \autoref{fig:all_error} shows error rates for all tasks collectively, with mean error rates per technique on the left and mean differences on the right. Mean error rates are lower for \Sym\ (7.4\%) followed by \Dor\ (8.1\%) and \Bar\ (8.6\%). There is no evidence that error rates were different across techniques. Thus the main differentiation we can make across techniques comes from completion time.\\

\begin{figure}[h]
 \vspace{-5pt}
    \includegraphics[width=\linewidth]{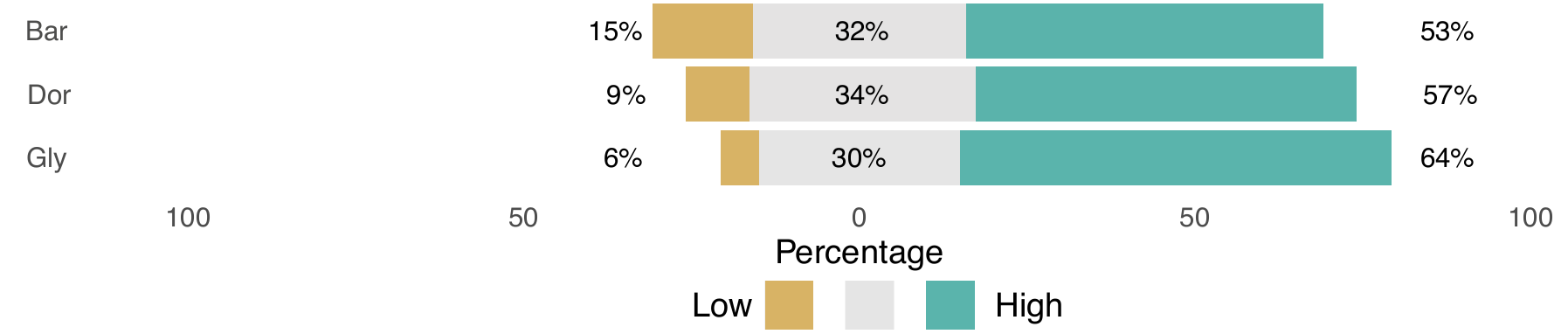} 
    \vspace{-20pt}
     \caption{Self-reported confidence across all tasks per visualization.
     }\label{fig:all_conf}
         \vspace{-5pt}
\end{figure}

\noindent \textbf{Confidence:} \autoref{fig:all_conf} shows the self-reported confidence for each visualization, for all tasks. Confidence is high for all three visualizations in more than half the trials, although more so for \Sym\ (64\% of trials) than for \Dor\ (57\%) and \Bar\ (53\%). 

\subsection{Results per task}
\begin{figure*}[h]
\begin{center}
 \vspace{-10pt}
\includegraphics[width=1\linewidth]{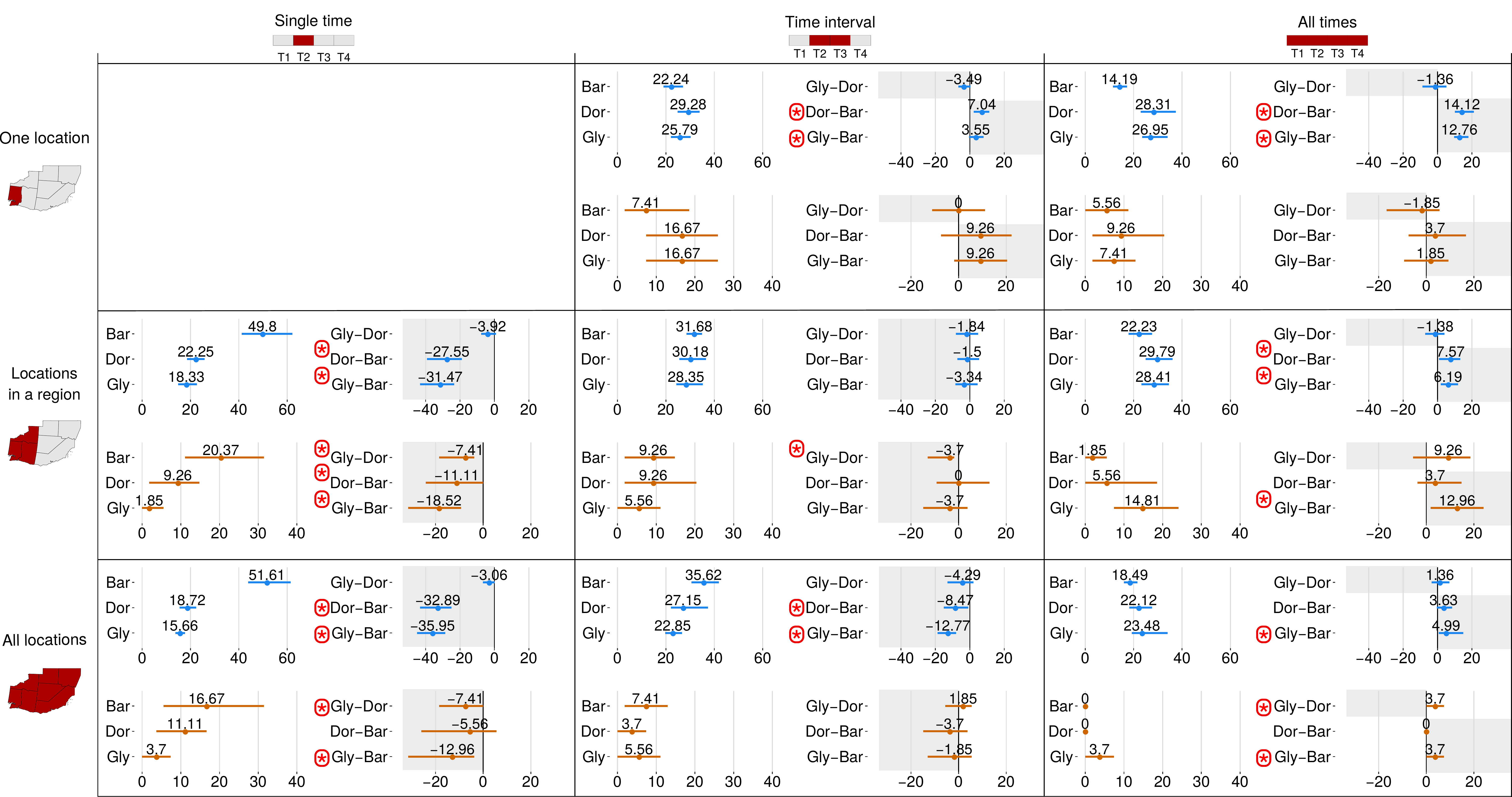}
    \end{center}
    \vspace{-15pt}
     \caption{Results for \textcolor{blueCT}{Completion Time}  (sec) and \textcolor{orangeER}{Error Rate}
     	(in \%) for each task \nc{in \autoref{fig:tasks}}.
		\nc{In each cell (task), Mean values per visualization are seen on the left and means of pairwise differences on the right}.  
		Error bars represent 95\% Bootstrap confidence intervals.  
		Gray rectangles indicate the direction of our hypotheses. 
		\nc{Evidence of differences are marked with a \protect\starEvidence\  (the further away from 0 and the tighter the CI, the stronger the evidence).}
		}
     \label{fig:summary}
\end{figure*}

Next we report results per task, grouped by \emph{temporal granularity} \nc{for legibility purposes (analyses were performed per task). The values and CIs for means and differences of means for both Completion Time and Error Rate can be seen separately for each task in \autoref{fig:summary}, with the direction of our hypothesis indicated by a gray background.} Self-reported Confidence per task can be seen collectively in \autoref{fig:confidence}.

\subsubsection{\nc{ \textsc{SingleTime} correlation tasks}}

\nc{In tasks involving a single time step we expect small multiples techniques to fare better (\textbf{H1}). Completion times and error rates (means and CIs) for these tasks are found in the leftmost column of \autoref{fig:summary}.}


{\textbf{\CT:} is faster with small multiples (\Sym\ , \Dor\ ) and slower for \Bar\ for both geographic granularity tasks.}
Looking at mean differences, there is strong evidence that \Bar\ is slower than both small multiples techniques, by more than 27sec for \tinyGeoReg\ \textsc{Region}, and by more than 32sec for \tinyGeoAll\ \textsc{AllLocations} tasks. Results are inconclusive for the difference between \Sym\ and \Dor\ in both tasks.

{\textbf{\ER:} Similar to the results for completion time, for both \tinyGeoReg{}~\textsc{Region} and \tinyGeoAll\ \textsc{AllLocations} tasks, \Sym\ had the best performance, followed by \Dor\ and \Bar\ with the highest error rate. Looking at mean differences, there is strong evidence that \Bar\ is more error prone than \Sym\ for both types of geographic granularities. There is weak evidence that \Bar\ is also more error-prone than \Dor\ for \tinyGeoReg\ \textsc{Region} (but no evidence of a difference for \vGeoA{}). Finally, \Dor\ appears more error-prone than \Sym\ for both tasks (strong evidence of this difference for \vGeoR{}, and weak for \tinyGeoAll\ \textsc{AllLocations}).

{\textbf{Confidence:} (self-reported by participants) corroborates these findings. For both tasks that considered \textsc{Single Time}, confidence is high for small multiples techniques (\Sym\ and \Dor) but low for \Bar\ (see top of \autoref{fig:confidence}).}

\paragraph*{Summary for {\sc SingleTime}:} 
Overall, the tendencies for the two tasks that focus on correlations for a {\sc Single Time} are similar, irrespective of whether we consider a geographical region or all locations. We have evidence that using the small multiples visualizations (\Sym, \Dor) takes less time (less than 20sec) and causes less errors than \Bar, supporting \textbf{H1}. There is also evidence of differences between \Sym\ and \Dor\ when it comes to errors, with \Dor\ being more error prone, supporting \textbf{H4}. 


\subsubsection{\nc {{\sc Time Interval} correlation tasks}}
\nc{In time interval tasks, we expect different performance across geographic granularities (\textbf{H2}), with a single map (\Bar) faring better for tasks involving one location (\textbf{H2.1}), and small multiples faring better for tasks involving a region or all locations (\textbf{H2.2}). Completion times and error rates (means and CIs) are found in the middle column of \autoref{fig:summary}.}


{\textbf{\CT:} When considering \vGeoL{}, we observe that completion time is indeed on average lower for \Bar\ (22.2sec), followed by \Sym\ (25.8sec) and then \Dor\ (29.3sec). Looking at the mean differences, there is evidence that  \Bar\ is faster than \Dor\ (by 7sec on average). It may be the case that \Bar\ is also faster than \Sym\ and that \Sym\ is faster than \Dor{}, but evidence is not conclusive.}

{The completion time for \vGeoR{} is close for all three techniques (\Sym\ 28.4sec, \Dor\ 30.2sec, and \Bar\ 31.7sec) and we do not have evidence of differences looking at the mean differences. The same pattern is found in \vGeoA{} as \Sym\ (22.8sec) is faster than the other techniques, followed by \Dor\ (27.1sec) and \Bar\ (35.6sec). Looking at the mean differences, we have evidence that \Bar\ is slower than both \Sym\  and \Dor\ (by 12.7sec and 8sec on average). We do not have evidence of a difference between \Sym\ and \Dor.}

{\textbf{\ER:} For these tasks, we observe that the lowest error rate depends on the geographical granularity considered. \Bar\ is better for \vGeoL{} (7.4\%), \Sym\ for \vGeoR{} (5.6\%) and \Dor\ for \vGeoA{} (3.7\%). Looking at mean differences for \vGeoA{} there is indeed evidence that \Dor\ is more error prone than \Sym\ (by 3.7\% on average) for \vGeoR{}, but no evidence of other differences.}

{\textbf{Confidence:} The second row of \autoref{fig:confidence} shows the self-reported confidence for \textsc{Time Interval}. We observe that confidence for \vGeoL{} is high in more than half of the trials for \Bar\ and \Sym\ (over 60\%), but lower for \Dor\ (45\%). For tasks in \vGeoR{} and \vGeoA{}, we observe that it is higher for both \Sym\ and \Dor\ (over 60\%) and lower for \Bar\ (54\% and 50\% respectively).}

\paragraph*{Summary for {\sc TimeInterval:}} The tendencies for the three tasks that focus on correlations for a time interval change significantly depending on the spatial granularity. For a single location, \Bar\ is faster than the small multiple techniques (\Sym,\Dor), supporting \textbf{H2.1}. 
 This behavior is reversed when considering all locations on the map. \Bar\ becomes the slowest visualization, supporting the part of \textbf{H2.2} related to all locations. 
 In both tasks, we found no evidence of difference in error rates.
 The situation is less clear when multiple locations in a region have to be considered. We found no evidence of differences for any of the measures, contrary to the prediction of \textbf{H2.2} related to geographical regions. 
 We observe no difference between \Sym{} and \Dor{}. \textbf{H4} is thus not supported.
 
\subsubsection{\nc{\textsc{All Time}}}

\nc{In tasks involving all time steps we expect a single map (\ie{}, \Bar) to fare better (\textbf{H3}). Completion times and error rates (means and CIs) for these tasks are in the rightmost column of \autoref{fig:summary}.}


{\textbf{\CT:} is lower with \Bar\ than with both small-multiples visualizations. Looking at the mean differences, there is strong evidence that \Bar\ is faster than \Sym\ and \Dor\ for both \vGeoL{} and \vGeoR{} tasks. For \vGeoA{} task, there is also strong evidence that \Bar\ is faster than \Sym\ (by 4.9sec on average) but evidence is not conclusive regarding \Bar\ being faster than \Dor. There is no evidence of a difference between \Sym\ and \Dor\ for any geographical granularity.}

{\textbf{\ER:} is lowest in \Bar\ for \vGeoL{} and \vGeoR{} tasks. For \vGeoA{}, the error rate is 0\% for both \Bar\ and \Dor{} (and thus, no CI is computed). There is evidence that \Bar\ is less prone to errors than \Sym\ for \vGeoR{}, but this evidence is weak for \vGeoA{} (and we see no evidence of a difference for \vGeoL{}). There is also weak evidence that \Dor\ is also less error prone than \Sym\ (by 3.7\%) for \vGeoA{}.}

{\textbf{Confidence:} is high in over 60\% of trials for most visualizations  and geographic granularities, with high-confidence trials for \Dor\ being a bit lower (around 50\% of trials) for the \vGeoL{} and \vGeoR{}. 

\paragraph*{Summary for {\sc AllTime}:} 
The tendencies for the three tasks that focus on correlations over all time steps are fairly similar, with \Bar\ being generally faster than small multiples (\Sym, \Dor), thus supporting  \textbf{H3}. Again, we do not find evidence of a difference between \Sym\ and \Dor{}. \textbf{H4} is not supported.

\begin{figure}[t]
 \vspace{-10pt}
 \includegraphics[width=\linewidth]{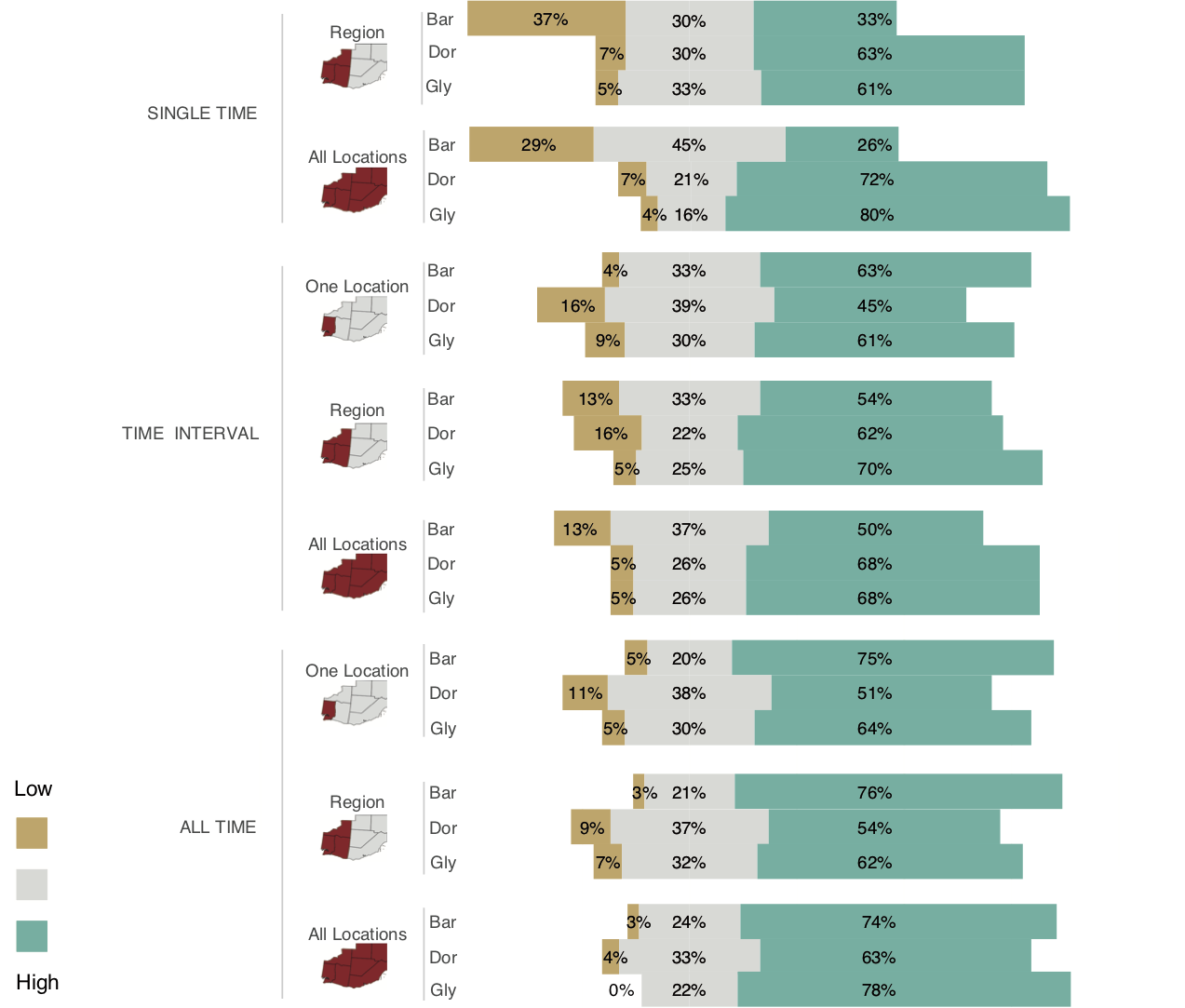} 
    \vspace{-20pt}
     \caption{Reported self-confidence per task (in \%).  
     }\label{fig:confidence}
\end{figure}

\section{Per-Task Discussion and Design recommendations} \label{sec:discussion}


We observed that, overall, small multiples were faster across tasks, but their error rates were not different from those of a map with bar charts. Nevertheless, as hypothesized, looking at the individual tasks we see that the performance changes depending on the task at hand. Next, we summarize and discuss our findings, and distill them into design recommendations (summarized in ~\autoref{fig:sum}).

\paragraph*{\sc SingleTime:} The correlation of thematic variables on geographic maps has been studied before for a single point in time~\cite{Gao2018-so, Elmer2013-vw}. We add to these findings, by identifying that the tendencies for correlation tasks on a single point in time are similar, irrespective of whether we consider a geographical region or all locations. Using the small multiple visualizations (\Sym{}, \Dor), participants were almost twice as fast as when using a single map with bar charts (\Bar), as they only needed to focus on a single cell of the small multiples, since that cell juxtaposes all spatial information for that point in time. 
The tasks are slower with a single map with bar charts (\Bar), since participants needed to visually search for the specific time step across multiple bar charts and synthesize their findings. 
Error rates for these tasks follow similar trends. Our findings thus confirm \textbf{H1}.

When it comes to small multiples, there is a tendency for the proportional symbol map  (\Sym{}) to be less error prone than the Dorling cartogram (\Dor), supporting \textbf{H4}. 
This is likely the case because the position of symbols shifts between multiples in the cartogram case, making it hard to re-identify them.
This tendency was also observed when comparing proportional symbol maps with 
non-contiguous cartograms in the work of Gao \etal~\cite{Gao2018-so}.
However, in their case, it was for the overall performance over multiple tasks, not just
for correlation identification, and the differences observed were not significant.

These tendencies were consistent with the self-perceived difficulty of conducting 
these tasks in the exit questionnaire.
It was stated often that it is hard to make analyses for 
one time step with bar charts.

\emph{\textbf{R1:} For identifying correlations at a specific point in time, small-multiples visualizations are better.}

\paragraph*{\sc TimeInterval:} When participants have to identify correlation tendencies and evolution over a time interval, the situation is less clear. The tendencies change significantly depending on the geographic granularity (consistent with \textbf{H2}).  When considering a single location, a single map with bar charts (\Bar) is faster than the small-multiples techniques (\Sym, \Dor), as all temporal information is grouped closely together and participants just needed to identify the temporal interval on a single bar chart. Whereas for small multiples, after identifying the relevant time cells, participants needed to then identify, in each cell, the specific location and collate their findings. This is consistent with \textbf{H2.1}. 

The findings are reversed when considering all locations, consistent with \textbf{H2.2}. Here, a map with bar charts is slower, because it is the visualization where information is scattered and needs to be collated from across different areas. Participants first had to go through all (or almost all) bar charts to identify the specific interval, and collate the information to identify tendencies. Whereas for small multiples, they only needed to focus on a few time steps and look for overall patterns.


\nc{One of the most interesting findings from this study is} the inconclusive evidence for tasks where a geographic region has to be considered across a time interval (this part of \textbf{H2.2} is not confirmed). The lack of observed differences may be due to low statistical power. But we believe it is more likely due to this task being more balanced in the amount of information that needs to be collated across different areas for the different techniques. Here, for a single map with bar charts, participants still had to identify the specific bars across multiple bar charts -- but not all of them. When using small multiples, they could focus on a few time steps, but still had to identify the desired geographic area in each one of them. There is likely a tradeoff when it comes to tasks that involve spatial regions and time intervals. When considering subsets of time, it looks like the less spatial locations have to be considered, the better a single map is. Inversely, the more spatial locations, the better small multiples become. \nc{More generally, it is likely that a single map with bar charts likely works best for simple geography and complex temporal patterns, and small multiples when geography is complex but the temporal variability is simple. Future work needs to determine exactly when to transition between visualizations.}
We are not aware of any previous work that has considered correlation tasks that require gathering information across subsets of space and time.

\emph{\textbf{R2:} For identifying correlations and temporal evolution over a subset of time steps and a subset of locations, there is no clear winner. If there are only a few locations, consider using a single map with bar charts. If there are many locations, prefer small multiples.}
 
\paragraph*{\sc All Time:} The tendencies for the three tasks that focus on correlations for all time steps are again consistent, with \Bar\ being faster, in accordance with \textbf{H3}. Even though participants had to collate both spatial and temporal information, a single map with bar charts was faster. This representation makes it easy to see  trends over time (correlation and monotonic evolution) that are juxtaposed in the individual bar charts. Collating this information seems to be fast irrespective of how many geographic regions are taken into account. Small multiples seem slower, likely because determining temporal trends necessitates comparing several locations across cells before identifying a trend. 

The self-perceived difficulty to conduct the task for all
time steps was also consistent with objective measures.
A single map with bar charts was perceived, overall, as easier to use 
than both small-multiples visualizations,
and 
 several participants commented that it was easy to observe 
evolution over time on the single map with bar charts.

\emph{\textbf{R3:} For identifying correlations and temporal evolution over all time steps, irrespective of the number of locations, a single map with bar charts is better.}

\paragraph{Small multiples:} We found evidence that the two small-multiple techniques
(\Sym, \Dor) were different mainly when considering a single point in time (partially confirming \textbf{H4}), with \Dor\ being slower and more error prone. 
Participants' comments 
indicate that they had difficulty matching a location, or sets of locations, across small multiples with \Dor, since positions of
locations shifted.
Nevertheless, this cost is not seen in tasks considering more than one time step. This may be due to low statistical power, or because this cost is small when it comes to more challenging tasks (time intervals or all time steps) that require collating information across small multiples.

\emph{ \textbf{R4:} For small multiples, there is some evidence that proportional symbol maps are better than Dorling, especially for a single time point.}

\begin{figure}[t]
\centering
 \includegraphics[width=0.7\linewidth]{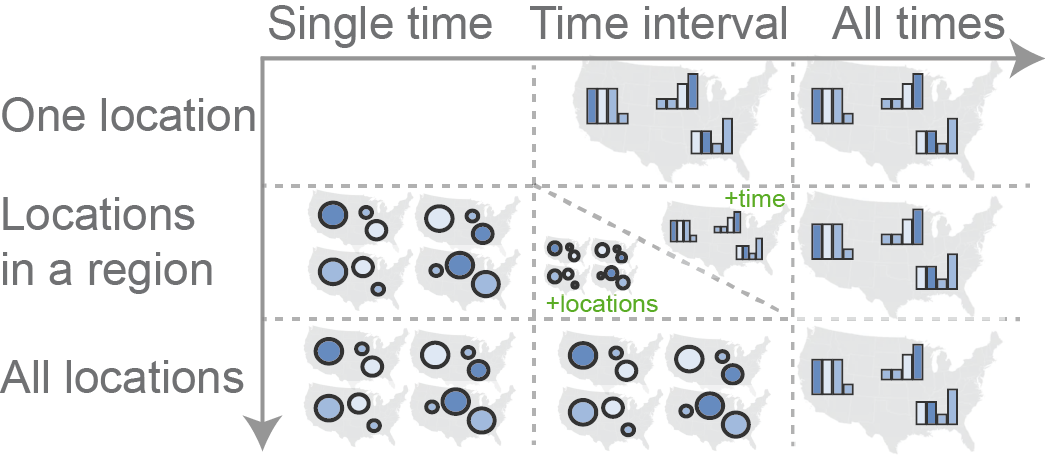}
 \vspace{-3pt}
\caption{\nc{Summary of our recommendations for the different tasks. For tasks on subranges of space and time (middle cell) there is no clear winner, but the table structure suggests small multiples work best for increased spatial complexity, and a single map with bar charts for increased temporal complexity (although the transition point is not known).} 
} \label{fig:sum}
\vspace{-15pt}
\end{figure}

%

\section{General Discussion}


Our findings generally followed our original hypothesis. We thus believe that our reasoning, that difficulty in each technique depends on whether the information to be collated is juxtaposed or distant, is sound; and that our results reflect true tendencies.  

The one exception relates to correlation tasks on subranges in time and space. We had originally thought that small-multiple variations would prevail in this situation, but we were unable to detect a trend. We believe that our setup of this task may reflect a similar difficulty in collating temporal information (for small multiples) and spatial information (for the single map with bar charts).  Looking at ~\autoref{fig:sum}, we observe that this middle cell is a transition cell between tasks better suited to small multiples and tasks better suited to single maps. For example, if we look at the middle column (Time interval), it seems that one (and likely a few) geographic locations are best seen on single maps, but as more locations are added small multiples become more competitive. Or if we look at the middle row (Locations in a region) it seems that one (and likely a few) time steps may be better viewed with small multiples, but as time increases a single map with bar charts becomes better. It is interesting to consider what are the tipping points of these shifts (number of time steps, number of geographic locations), in order to determine when to transition between visualizations. 


For all visualizations, collating information across different areas (bars from different bar charts for the single map, and locations across cells for the small multiples) is challenging. In our study, we focus on static visualizations, but the addition of highlighting would likely reduce the differences we found, by making it easier to collate information (\eg{}, highlight Washington in all small multiples, or 2011 in all bar charts). Nevertheless, we believe the high-level effects would still hold (to a lesser extent) as they are due to the fact that information is dispersed across the visualization. 
If filtering is considered, we believe behavior will likely revert to the results at the corners of ~\autoref{fig:sum}. For example, filtering on time interval 2009-2011 would either remove or fade other years out, making this a task similar to ones over all time steps. Similarly, if the East US is the focus, the system would remove or fade other locations out, making this a task similar to those involving all locations. 
More importantly, the actions performed to select or filter time steps or geographical locations could themselves be used as an indication of what is the user's focus, and used to transition to the best visualization for the task. 


%

%

\subsection{Limitations and Future Work}
\label{sec:limfw}

Interaction was deliberately not considered in this first study, as we primarily aimed at evaluating the specific influence of space and time at different granularities on users' ability to identify correlations with different visualizations. Thus, we wanted to avoid adding further factors to an already complex experimental design. Our discussion section \nc{above} provides initial thoughts about how interaction could affect our results, but further work is needed to verify them, and to consider the use of interaction as a means to transition between visualizations.

The number of steps used to detect correlation in our tasks is limited 
(nine time steps per location), which required us to use datasets with a strong 
relationship between two variables.
Data extracted from measures of real world phenomena is unlikely to present 
such strong patterns, making it harder to detect potential correlations. 
This could alter 
our results, although we believe the general trends would persist.

Another limitation is that we only used a single map of the US, which necessarily captures only a subset of geographical configurations.
\nc{It is possible that countries with more diverse shapes (\eg\ Chile, Italy 
or United Kingdom) would lead to different results, as the identification of 
individual locations or regions might be different.
We attempted to mitigate any bias in identifying the locations of interest by displaying, prior to each trial, the 
geographic region of interest. Nevertheless, further experimentation is needed. 
Moreover, diversity of irregularity of locations can impact spatial autocorrelation in geospatial visualizations that use 
irregular geometries to represent thematic variables, such as choropleths~\cite{Beecham2017-bi,mcnabb2018size, ward2015interactive}. While it is possible that effects might differ somewhat in other types of maps, we feel that the general trends should hold: our techniques use regular shapes to 
represent thematic variables, and thus the size and number of items compared likely weigh more in the complexity of the task (\eg{}, occlusion or clutter of elements might impact 
the interpretation of patterns). To this end, in our trials we varied the size of locations and their density. 
Finally, although the analysis of data using a map of a known country could 
have led to bias given preconceptions about the geographical distribution,
we believe this to be unlikely given the extensive training, and the 
number of map features and time steps involved.\footnote{ \nc{We did not find any warning signs of such pitfalls (\eg{}, participants
taking very little time to finish the tasks and making numerous errors).}
}
In summary, while we believe that overall trends would persist across different maps, future work needs to consider more diverse geographic
maps.}

\nc{For the small multiples tested, we expected that Dorling cartograms  (\ie , visualizations that use visual channels of the maps features themselves to encode thematic variables) would perform worst than proportional symbol maps,
as was the case in previous work~\cite{Gao2018-so, Kaspar2011-rb}. 
In our context this was observed mainly when considering tasks at a single point in time. It is very likely this effect will be more pronounced in other spatial tasks that involve more continuous geographic changes and correlations that vary spatially (\eg{}, identifying transmission patterns).} 

We recruited users who were already knowledgeable about visualization, and gave them additional training.
Opportunities for such training may not be available to the general public.
While we believe general trends will still apply,
\nc{it is possible that non-trained users would have lower accuracy rates
or would not dedicate as much time as our participants to perform the
tasks.
Additionally, they might be more familiar with one of the three tested 
techniques, which would bias results in its favor.}
Future work should investigate the learning curve of each 
technique and analyze how well they fare when used by novices with a more diverse background and lack of training.
A next step in that direction would be to conduct a crowdsourced study.

Finally, we decided to combine two different association tasks in one (\ie{}, the type of correlation and its evolution), as we felt they were tightly coupled when performed in the context of geo-temporal analysis.
Due to this combination, our analysis does not provide finer details on the difficulty of each subtask.
Future work could study each one separately to gain more
insights about how correlation and trends are detected individually. For example, we expect that \nc{complex temporal tasks, such as} detecting \nc{and characterizing} monotonic evolution, is easier on single maps with bar charts (as each one directly encodes this evolution); \nc{whereas complex geographical tasks, such as detecting transmission patterns, may be easier with small multiples.}


\section{Conclusions} \label{sec:conclusions}

We presented a study on identifying correlation in spatio-temporal 
visualizations. We considered eight tasks that associate two 
variables over different granularity levels for both time and space.
The compared visualizations \nc{combine different strategies}
to represent thematic variables: juxtaposing either time or space (a single map with bar charts {\em vs.} small-multiple maps); \nc{encoding variables either using symbols overlaid on top of map features, or using visual channels of the map features themselves} (proportional symbol maps {\em vs.} cartograms).
We provide a set of design recommendations depending on the task at hand.
In our context, \nc{the technique using the map features' own visual channels to encode thematic variables (cartograms)} 
performed worst only when a single point in time was considered. 
Our results further indicate that for tasks that consider the
evolution over all time steps, a visualization that represents data on a single map (juxtaposing time) is
more effective and easier to interpret than small multiples.
Small multiples (juxtaposing space) are better suited for tasks that require the
comparison of variables for one point in time over several geographical 
locations.
When dealing with time intervals and spatial regions, our results suggest that there is a continuum of performance between visual representations (juxtapose time \textit{vs.} space), raising questions for future research.



\bibliographystyle{abbrv-doi-hyperref}

\bibliography{ms}
\end{document}